\documentclass[twocolumn,prl,showpacs]{revtex4}
\usepackage{psfrag}
\usepackage{graphicx}
\usepackage{amsmath}
\usepackage{amssymb}

\begin{document}
\title{Inelastic Scattering in Metal-H$_2$-Metal Junctions}
\author{I. S. Kristensen$^1$, M. Paulsson$^2$, K. S. Thygesen$^1$, and
K. W. Jacobsen$^1$}
\affiliation{$^1$Center for Atomic-scale Materials Design (CAMD), \\
Department of Physics, Technical University of Denmark, DK - 2800 Kgs. Lyngby, Denmark}
\affiliation{$^2$Division of Physics, Department of Natural Sciences, Kalmar University, 391 82 Kalmar, Sweden
}
\date{\today}

\begin{abstract}
  We present first-principles calculations of the $dI/dV$
  characteristics of an $\text{H}_2$ molecule sandwiched between Au
  and Pt electrodes in the presence of electron-phonon interactions.
  The conductance is found to decrease by a few percentage at
  threshold voltages corresponding to the excitation energy of
  longitudinal vibrations of the $\text{H}_2$ molecule. In the case of
  Pt electrodes, the transverse vibrations can mediate transport
  through otherwise non-transmitting Pt $d$-channels leading to an
  \emph{increase} in the differential conductance even though the
  hydrogen junction is characterized predominately by a single almost
  fully open transport channel. In the case of Au, the transverse
  modes do not affect the dI/dV because the Au $d$-states are too far
  below the Fermi level. A simple explanation of the first-principles
  results is given using scattering theory. Finally, we compare and
  discuss our results in relation to experimental data.

\end{abstract}
\maketitle
In recent years it has become possible to measure the electrical
properties of single molecules captured between metallic
electrodes~\cite{venkataraman,kubatkin,smit02}. Such experiments
provide a unique opportunity to develop our understanding of basic
quantum mechanical phenomena at the nanometer length scale and at the
same time constitute the first steps towards molecule-based 
electronics.~\cite{introducing_molel}

Interactions between the conduction electrons and the molecule's
vibrational degrees of freedom is of particular interest for the
performance of molecular electronics devices as they determine the
local temperature and stability of the device when subject to an
external bias voltage\cite{diventra_nature}.
Moreover, inelastic scattering can be used to identify the
atomic structure of molecular junctions by exploiting the sensitiveness of
the molecule's vibrational frequencies and the electron-phonon 
interaction to the
junction geometry.~\cite{djukic,lorente,frederiksen3,wolf,hipps,ho99,galperin}

\begin{figure}[!t]
\includegraphics[width=1.0\linewidth]{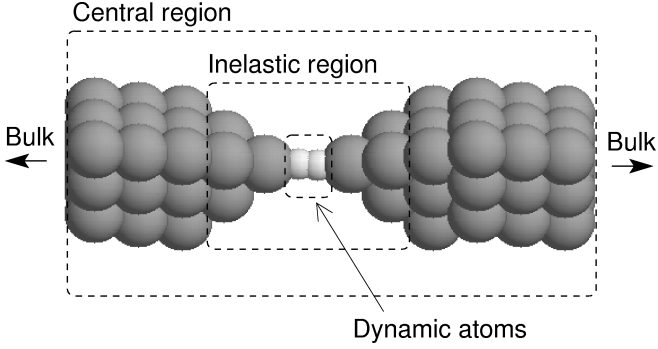} 
\caption{The supercell use to model the metal-$\textrm{H}_2$-metal
  junction. Only the hydrogen atoms are allowed to vibrate (the
  "dynamic" atoms). This is a good approximation due to the large
  difference in mass between Au/Pt and H. The effect of the field
  generated by the vibrating H atoms is taken into account inside the
  indicated inelastic region. The central region, $C$, is coupled to
  semi-infinite bulk electrodes and periodic boundary conditions are
  imposed in the directions perpendicular to the contact axis. }
\label{fig1}
\end{figure}

Perhaps the simplest molecular junction consists of a single hydrogen
molecule sandwiched between metal electrodes, see Fig.
\ref{fig1}.~\cite{smit02,halbritter} Shot noise measurements on 
Pt-$\textrm{D}_2$
contacts show that the conductance is carried predominantly by a
single almost fully transparent channel\cite{shotnoise}, and density
functional theory (DFT) calculations have shown that this is
consistent with a linear bridge
configuration. \cite{smit02,cuevas,thygesen_h2prl,prb_ptpd_h2}. An alternative
configuration where the H$_2$ molecule is dissociated in the contact has
also been proposed, however, this junction yields a conductance larger
than $1G_0$ ($G_0=2e^2/h$ is the conductance quantum) with
contributions from three channels~\cite{garcia}. Inelastic point
contact spectroscopy provides information about the hydrogen molecule's
vibrational frequencies and their variation upon stretching. The data
obtained from such measurements have also been found to be
consistent with the linear bridge configuration.\cite{djukic}

The fact that the hydrogen junction supports a single, almost fully
open conductance eigenchannel suggests that the inelastic scattering
processeses should be particularly simple to understand. Indeed, consider
a junction supporting a single scattering channel at the Fermi energy
with a transmission
probability of $\mathcal T=|t(\varepsilon_F)|^2$. At low
temperatures the molecule sits in its vibrational groundstate
and the electron looses the energy $\hbar \Omega$ to the molecule during a
scattering event. Assuming a bias voltage $eV=\mu_L-\mu_R>\hbar
\Omega$ an electron incident on the
molecule from the left with an energy just below $\mu_L$, must end up in a
left moving scattering state after interacting with the molecule. This 
follows from energy conservation and the Pauli principle.
Upon inelastic scattering, the probability for the electron to enter the
right electrode is thus changed from $\mathcal T$ to $\mathcal
R=1-\mathcal T$. Consequently, the change in conductance due to the
electron-phonon interaction should be proportional to 2$\mathcal T-1$, 
i.e. an increase
(decrease) in the conductance is expected for $\mathcal T<0.5$
($\mathcal T>0.5$). The same conclusion has been reached using more
rigorous arguments~\cite{delavega,paulsson05,viljas} and has recently
been supported by measurements on Pt-H$_2$O junctions~\cite{tal_h2o}.

In this paper we present DFT calculations for the $dI/dV$ curves of
Pt-H$_2$-Pt and Au-H$_2$-Au junctions in the presence of
electron-phonon interactions.  For both Pt and Au electrodes, scattering
on the longitudinal modes lowers the conductance by a few percentage of
$G_0$ in accordance with the simple one-channel model discussed above. In 
the case of Pt, the
transverse modes can mediate tunneling through the otherwise closed
$d$-channels leading to an \emph{increase} in the conductance of up to
$5\%$ of $G_0$, demonstrating that the metal-H$_2$-metal junction cannot be
viewed as a simple one-channel system. For Au, the transverse modes have no
effect on the conductance because only $s$-states are present at the
Fermi level and these do not couple via the transverse vibrations.

The Hamiltonian of the system is given by
\begin{equation}
\hat H=\hat H_{\textrm{el}}+\hat H_{\textrm{ph}}+\hat H_{\textrm{el-ph}},
\end{equation}
where $\hat H_{\textrm{el}}$ is the Hamiltonian of electrons moving in the
\emph{static} equilibrium structure, $\hat H_{\textrm{ph}}$ describes the
vibrations of the $\textrm{H}_2$ molecule, and $\hat H_{\textrm{el-ph}}$ is the
interaction between the electrons and the vibrating hydrogen atoms.
For $\hat H_{\textrm{el}}$ we use the Kohn-Sham Hamiltonian.

Within the harmonic approximation the molecular vibrations are
described by the Hamiltonian 
$\hat H_{\textrm{ph}}=\sum_{\lambda}\hbar \Omega_{\lambda}
(b^{\dagger}_{\lambda}b_{\lambda}+\frac{1}{2})$
where $b^{\dagger}_{\lambda}$ ($b_{\lambda}$) creates (destroys) a
phonon in mode $\lambda$. The electron-phonon interaction takes the form
\begin{equation}\label{eq.elph}
\hat H_{\textrm{el-ph}}=\sum_{n,m\in C}\sum_{\lambda} M^{\lambda}_{nm}c^{\dagger}_n c_m(b^{\dagger}_{\lambda}+b_{\lambda})
\end{equation}
where the first sum runs over Wannier functions located in the
inelastic region, see Fig. \ref{fig1}, and the second sum runs over
vibrational modes. The electron-phonon coupling matrix, $M^{\lambda}$, is given 
by $M^{\lambda}_{nm}=\langle \phi_n(\bold
r)|W^\lambda(\bold r)|\phi_m(\bold r)\rangle$, where the displacement 
potential, $W^\lambda(\bold r)=\nabla v_s[\{\bold
R_{n}\}](\bold r)\cdot \bold Q_{\lambda}$, is the derivative of the
effective KS potential in the direction defined by eigenmode
$\lambda$. In practice $W^{\lambda}$ is obtained as a finite difference 
between equilibrium Hamiltonians describing the electronic system
when the hydrogen molecule has been moved in the positive and the negative 
normal direction.

The current flowing into the molecule (central region $C$) from lead 
$\alpha=L,R$ is calculated from the formula~\cite{haug_jauho,meir_wingreen}
\begin{equation}\label{eq.current}
I_{\alpha}=\frac{e}{h}\int \textrm{Tr}\Big[ \normalsize \Sigma_{\alpha}^{<}(\varepsilon)G_C^{>}(\varepsilon)-\Sigma_{\alpha}^{>}(\varepsilon)G_C^{<}(\varepsilon)\Big]
\normalsize
  \textrm{d}\varepsilon
\end{equation}
where $G_C^{<,>}$ is the lesser and greater Green functions of the
central region evaluated in the presence of coupling to
leads and the phonons~\cite{voltage}. 

The lesser and greater Green functions are given by
\begin{equation}\label{eq.ggl}
G^{\lessgtr}(\varepsilon)=G^r(\varepsilon)\big[\Sigma_{L}^{\lessgtr}(\varepsilon)+\Sigma_{R}^{\lessgtr}(\varepsilon)+\Sigma_{\textrm{ph}}^{\lessgtr}(\varepsilon)\big] G^a(\varepsilon)
\end{equation}
where $G^r(\varepsilon)=[\varepsilon+i\eta-[H_{\textrm{el}}]_{C}-\Sigma_{L}^r-
\Sigma_{R}^r-\Sigma_{\textrm{ph}}^r]^{-1}$ and 
$G^r(\varepsilon)=[G^a(\varepsilon)]^{\dagger}$.

The self-energy originating from the coupling to the leads are calculated using 
standard techniques~\cite{thygesen_chem_phys05}. For the self-energy due to the 
electron-phonon coupling from mode $\lambda$ we use the first Born 
approximation, 

\begin{eqnarray}\label{sel.ph}
\Sigma_{\textrm{ph},\lambda}^{\lessgtr}(\varepsilon)&=&M^{\lambda}G_0^{\lessgtr}(\varepsilon \pm\hbar\Omega_{\lambda})M^{\lambda}\\\nonumber \Sigma_{\textrm{ph},\lambda}^{r}(\varepsilon)&=&\frac{1}{2}\big[\Sigma_{\textrm{ph},\lambda}^{>}(\varepsilon)-\Sigma_{\textrm{ph},\lambda}^{<}(\varepsilon)\big]\\\label{ser.ph}
&-&\frac{i}{2}\int \frac{\Sigma_{\textrm{ph},\lambda}^{>}(\varepsilon')-\Sigma_{\textrm{ph},\lambda}^{<}(\varepsilon')}{\varepsilon-\varepsilon'}d\varepsilon',
\end{eqnarray}
where the last equation follows from the general identity
$G^r-G^a=G^>-G^<$ together with the Kramer's Kronig relation between
$\text{Im}\Sigma^r$ and $\text{Re}\Sigma^r$.  We assume zero phonon
temperature corresponding to infinite cooling of the vibrations, and thus the
number of phonons has been set to zero. Consequently electrons
never interact with an excited molecule and therefore can only lose
energy to the molecule during a scattering event.

As done often we have omitted the Hartree term
in the electron-phonon self-energy.\cite{frederiksen,gagliardi} The corresponding energy-independent contribution to 
the retarded self-energy can be 
understood as a static phonon-induced change in the mean-field electronic 
potential. It is expected that this small static potential
 would
be, at least partially, screened if included in the DFT self-consistency loop.


The supercell geometry of the considered hydrogen contact is shown in
Fig.~\ref{fig1}. The
distance between the two electrodes, or equivalently the length of the
supercell, has for the case of Pt been chosen to make the calculated vibrational
frequencies of the H$_2$ molecule match the experimental values as
close as possible\cite{djukic}. For the case of Au where less detailed experimental data is available, we have chosen the distance by minimizing the toal energy. Using the plane-wave
pseudopotential code Dacapo\cite{dacapo} we have relaxed the surface layers, 
the pyramids and the
hydrogen molecule to obtain stable junction structures. We used an energy 
cut-off of
25 Ry for the plane-wave explansion, described the ion cores by
ultrasoft pseudopotentials\cite{vanderbilt90}, and used a $1\times 4
\times 4$ Monckhorst pack grid for the $k$-point sampling. Exchange
and correlation effects were described with the PW91 functional.\cite{pw91} 
As a basis for the electronic states we use partially occupied maximally localized Wannier functions\cite{wf} which
allows for an efficient and accurate calculation of transport
properties as described in Ref.~\onlinecite{thygesen_chem_phys05}.

\begin{figure}[!t]
\includegraphics[width=1.0\linewidth]{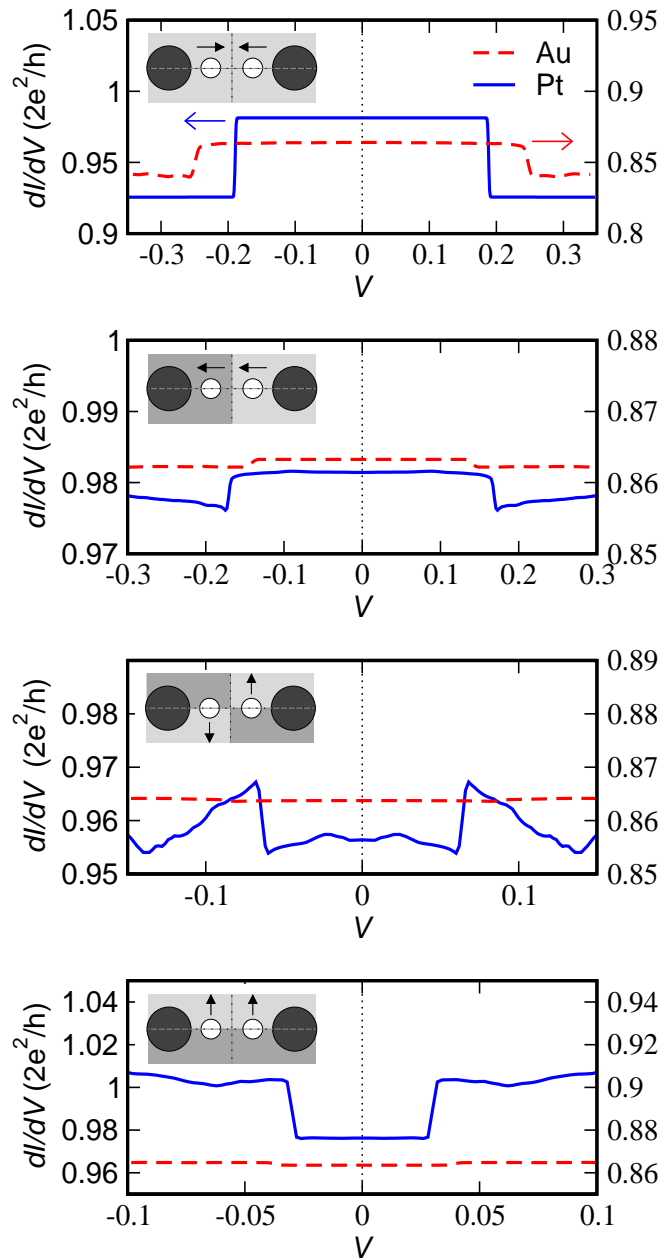} 
\caption{(Color online)Differential conductance of the Pt-$\textrm{H}_2$-Pt 
  (full) and
  Au-$\textrm{H}_2$-Au (dashed) junctions when scattering on a single
  vibrational mode is included. The insets illustrate the
  vibrational modes together with the symmetry
  of the corresponding displacement potential $W^\lambda(\bold
  r)$. Frequencies of the H$_2$ vibrational modes (in meV) for
Pt: $\hbar \Omega_{M1}= 190$, $\hbar \Omega_{M2}=171$, $\hbar \Omega_{M3}=64$,
$\hbar \Omega_{M4}= 30$ and for Au: $\hbar
\Omega_{M1}=249$, $\hbar \Omega_{M2}=141$, $\hbar \Omega_{M3}=84$, $\hbar
\Omega_{M4}=37$.}
\label{fig2}
\end{figure}

The vibrational eigenmodes, $\{{\bf Q}_{\lambda}\}$, and corresponding
frequencies, $\{\Omega_{\lambda}\}$, of the $\textrm{H}_2$ molecule are
obtained by diagonalizing the dynamical matrix of the system which in
turn is calculated from the DFT total energies by finite differences. Thanks to
the large difference in mass between the metal and hydrogen atoms, we
can calculate the dynamical matrix for the two H atoms keeping all
metal atoms fixed. Following this procedure we obtain a longitudinal
stretching mode (M1), a longitudinal center-of-mass mode (M2), as well
as two pairwise degenerate transverse modes which we refer to as
hindered rotations (M3) and hindered transverse center of mass modes (M4). The
modes are sketched in the insets of Fig.~\ref{fig2} and the corresponding 
frequencies are given in the caption. 

In Fig.~\ref{fig2} we show the differential conductance calculated
from Eq.~(\ref{eq.current}) including scattering on the different
vibrational modes separately. To extract the features due to the
inelastic scattering from those due to elastic scattering we have subtracted the
elastic signal, i.e. we plot 
$G(V)=G_{\textrm{full}}(V)-G_{\textrm{el}}(V)+G_{\textrm{el}}(V=0)$, see
Ref. \onlinecite{frederiksen} for a discussion of this procedure.

The conductance curves of Fig. \ref{fig2} present several interesting
features: For both Pt and Au the longitudinal modes lead to a
decrease in the conductance as expected from the one-channel
model. It is noticed that the internal stretching mode has a much larger
impact on the electrons than the CM mode. For Au, the transverse modes
have no effect on the transport, while for Pt they lead to an \emph{increase}
in the conductance. Since the junction has one fully open channel this
seems to conflict with the one-channel model which would predict an
increase only for junctions with conductance $<0.5G_0$. It is noted that
the differences in the zero-bias conductances are due to the tails of the
electron-phonon self-energy, which although centered around the vibrational 
frequencies also have weight at other energies. Before discussing the origin of
the above mentioned features it is useful to consider a simplified 
description of the scattering process.

In the following we regard $\hat H_{\textrm{el-ph}}$ as a perturbation to $\hat H_0=\hat
H_{\textrm{el}}+\hat H_{\textrm{ph}}$ and consider the scattering of a single electron
off a molecule in its vibrational groundstate. For
simplicity we disregard the effect of all the other
electrons (but we do take the Pauli principle into account). The scattering states of $\hat H_0$ are
conveniently chosen as the eigenchannels incident on the molecule from
the left, $\psi_{Lp}(\varepsilon)$, or right,
$\psi_{Rq}(\varepsilon)$.\cite{taylor,brandbyge_eig} The
probability that an electron of energy $\varepsilon$ injected from the
left lead in mode $p$, is transmitted (reflected) upon scattering
\emph{elastically} on the central region is denoted by
$\mathcal T_{Lp}(\varepsilon)=|t_{Lp}(\varepsilon)|^2$
($\mathcal R_{Lp}(\varepsilon)=|r_{Lp}(\varepsilon)|^2$). Due to the
non-mixing property of the eigenchannels we have $\mathcal
T_{\alpha p}+\mathcal R_{\alpha p}=1$ for all channels $p$ and $\alpha=L,R$. In terms of the eigenchannels the
Landauer formula for conductance takes the form 
$G_{\textrm{el}}=G_0\sum_{p}\mathcal T_{L
  p}(\varepsilon_F)=G_0\sum_{q}\mathcal T_{R q}(\varepsilon_F)$. The
state of the molecule is specified by the number of phonons in each
mode, $|\bold n\rangle$. We use the symbol $\Psi$ to denote a state of the combined electron-molecule system.

Assume that $eV=\mu_L-\mu_R>0$ and consider an electron incident
on the junction from the left in the state
$\psi^{\textrm{in}}=\psi_{Lp}(\varepsilon)$ with
$\mu_R<\varepsilon<\mu_L$ and the molecule in its vibrational
groundstate, $|\bold 0\rangle$. 
According to scattering theory, the system
ends up in the asymptotic out state, $\Psi^{\textrm{out}}=\hat S |Lp; \bold 0\rangle$, where $\hat S$ is the
scattering operator incorporating the effect of $\hat H_{\textrm{el-ph}}$. 
In the first Born approximation we have the transition amplitudes
\begin{eqnarray}\nonumber
\langle \alpha q; \bold n|\hat S| Lp; \bold
0 \rangle \approx
\langle \alpha q; \bold n | Lp; \bold 0 \rangle \\ - 2\pi i\delta(E_{\textrm{in}}
-E_{\textrm{out}})\langle \alpha q; \bold n |\hat H_{\textrm{el-ph}}|Lp; \bold
0 \rangle,
\end{eqnarray}
where $E_{\textrm{in}}$ and $E_{\textrm{out}}$ are the total energies of
the combined electron phonon system in the in- and out-going
states. This allows us to express the out state as
\begin{equation}\label{eq.finalstate}
  \Psi^{\textrm{out}}=C_p\Big [\psi_{Lp}(\varepsilon)\otimes
  |\bold 0\rangle+{\sum_{q,\lambda}}^\prime
  c_{pq}^\lambda\psi_{Rq}(\varepsilon-\hbar \Omega_\lambda)\otimes
  |1_\lambda\rangle \Big ]
\end{equation}
where the prime in the sum means that only modes with $\hbar
\Omega_\lambda<eV$ are included. The expansion coefficients are 
\begin{equation}\label{eq.expansion}
c_{pq}^\lambda= D_{Rq}(\varepsilon-\hbar
\Omega_\lambda)\langle \psi_{Rq}(\varepsilon-\hbar\Omega_\lambda)|W^\lambda (\bold r)|\psi_{Lp}(\varepsilon)\rangle
\end{equation}
where $D_{Rq}(\varepsilon)$ is the electronic density of states for
channel $Rq$. The normalization constant,
$C_p=(1+\sum_{q,\lambda}'|c_{pq}^\lambda|^2)^{-1/2}$, has been
introduced because the first Born approximation is not a unitary
approximation to $\hat S$. The fact that only states coming from the
right electrode are included in the sum of Eq. (\ref{eq.finalstate})
is a simple consequence of the Pauli principle.

In the elastic case, an electronic wavepacket constructed from the
states $\psi_{Lp}$ in a narrow interval around the energy
$\varepsilon$, initially located far from the molecule in the left
lead, will make it to the right lead with probability $\mathcal
T_{Lp}(\varepsilon)$. In contrast the scattered state
(\ref{eq.finalstate}) describes a situation where the initial
wavepacket makes it to the right electrode with probability
\begin{equation}
P_p = |C_p|^2\Big [\mathcal T_{Lp}(\varepsilon_F)+{\sum_{q,\lambda}}^\prime
|c_{pq}^\lambda|^2\mathcal R_{Rq}(\varepsilon_F)\Big ],
\end{equation}
where we have assumed that $\mathcal T$ and $\mathcal R$ varies little
on the scale of $\hbar \Omega$. The total change in conductance due to
the inelastic scattering can then be obtained from Landauer's formula
\begin{equation}\label{eq.change}
\Delta G = G_0\sum_{p}|C_p|^2{\sum_{q,\lambda}}^\prime
|c_{pq}^\lambda|^2\Big [\mathcal R_{Rq}(\varepsilon_F)-\mathcal T_{Lp}(\varepsilon_F) \Big ]
\end{equation}

Apart from the assumptions of instantaneous cooling of the phonons and
weak electron-phonon interaction, which also underlie the
first-principles results, Eq. (\ref{eq.change}) was derived in the
absence of a Fermi sea. However, as we show below,
Eq. (\ref{eq.change}) provides a simple and physically appealing
explanation of the
first-principles results of Fig. (\ref{fig2}).

It follows from Eq. (\ref{eq.change}) that the change in conductance involves 
all pairs of channels for
which the matrix element $\langle \psi_{Lp}|W^\lambda (\bold
r)|\psi_{Rq}\rangle$ is non-zero for some mode $\lambda$. Since
$W^\lambda(\bold r)$ extends to the metal atoms binding to H$_2$, any 
scattering state
-- transmitting or not -- with weight on these atoms will also contribute in
Eq. (\ref{eq.change}). 

In the case of Pt, we find at the Fermi level two types of eigenchannels with 
sufficient weight on the hydrogen atoms and the contacting Pt atoms that the 
coupling 
matrix element will be significant. One eigenchannel is the almost fully open
$s$-channel and the others have $d$ character and very low transmission at $\varepsilon_F$. Since Au has no $d$-states at the Fermi level, only the
$s$-channel makes a contribution in Eq. (\ref{eq.change}).

For the longitudinal modes, M1 and M2, the symmetry of
$W^\lambda$ implies that \mbox{$s$-$s$} transitions are possible, but not \mbox{$s$-$d$}
transitions (\mbox{$d$-$d$} transitions are not excluded by symmetry, but
because of the vanishing overlap between $\psi_{Ld}$ and $\psi_{Rd}$).
Since $\mathcal R_s-\mathcal T_s\approx -1$ we should expect a drop
in conductance in agreement with the first-principles
calculations. On the hydrogen molecule, the $s$-channel has mainly
character of the $\textrm{H}_2$ anti-bonding orbital. This implies that
the product $\psi_{Ls}(\bold r)^*\psi_{Rs}(\bold r)$ is unchanged upon
reflection in the plane cutting through the H-H bond perpendicular to
the molecular axis. On the other
hand the potential $W^{M2}(\bold r)$ changes sign upon this reflection. Consequently, the matrix element $\langle \psi_{Rs}|W^{M2}|\psi_{Ls}\rangle$ will be almost zero, and this explains the weak signal observed for M2 as compared to M1.

The spatial shape of the $d$-states implies that
coupling to the $s$-channel is possible only via the transverse modes
M3 and M4, see the symmetry of $W^\lambda$ in the insets of Fig. \ref{fig2}. Limiting 
the sums in
Eq. (\ref{eq.change}) to these two relevant states we see that $\Delta
G$ becomes proportional to $\mathcal R_d-\mathcal T_s$. The increase
in conductance found for the transverse modes in the Pt contact can thus be explained
by a higher reflection probability of the low-transmitting $d$-channel
as compared to the transmission probability of the high-transmitting
$s$-channel. We stress that small changes in the transmission
probabilities for the $s$- or $d$-channels could change the sign of
$\Delta G$. The symmetry of the displacement
potential, $W^\lambda$, for the transverse modes prevents coupling
between two states with $s$-symmetry, which explains why the transverse modes 
do not affect the conductance of the Au junction. 

We notice that the calculated increase in conductance due to the 
transverse modes is not in agreement with the experimental data from 
inelastic point contact
spectroscopy for Pt-H$_2$-Pt junctions which show a conductance decrease.
Some of the possible explanations for this disagreement is:
\begin{itemize}
\item[-] According to Eq. (\ref{eq.change}), the size (and the sign)
  of $\Delta G$ is determined by the relative magnitude of the $s$-
  and $d$-channel transmissions. Even small changes here could change
  the sign of $\Delta G$. In this sense, the fact that we obtain an
  increase in conductance while experimentally a decrease is
  observed, should be viewed as a quantitative rather than a
  qualitative difference.
\item[-] In principle the 1BA applies in the limit of weak
  electron-phonon interactions while we obtain electron-phonon matrix
  elements ($M$ in Eq. (\ref{eq.elph})) on the order of electron
  volts. On the other hand the inelastic features in the dI/dV are a
  few percentage of $G_0$ indicating that only a few out of a hundred
  electrons are scattered. Moreover, previous studies applying the 1BA
  to gold chains agree nicely with experiments\cite{frederiksen3},
  indicating that the 1BA provides an accurate description of
  electron-phonon interactions in strongly coupled
  metal-molecule-metal junctions.
\item[-] The highly symmetric geometry of the metal-H$_2$-metal
  junction used in this study is an idealised but oversimplified model
  of the real structure. However, we have considered other less
  symmetric configurations none of which gave rise to a conductance
  decrease for the transverse modes.
\item[-] Inclusion of a finite phonon temperature could affect the
  calculated properties. However, as can be seen from Eq. (14) of Ref.
  \cite{paulsson06}, to lowest order in the electron-phonon
  interaction strength the sign of $\Delta G$ cannot change by
  including heating.
\end{itemize}

Despite the differences between the experimental and theoretical
findings for the phonon induced features in the $dI/dV$, we hesitate
to conclude that the linear bridge configuration is not the structure
observed in the experiments. The reason is the strong evidence
mentioned in the introduction which favors the linear bridge combined
with the small size and high sensitivity of the inelastic features.
 
In conclusion, we have performed first-principles calculations for the
non-linear $dI/dV$ curves of Pt-H$_2$-Pt and Au-H$_2$-Au molecular
junctions in the presence of electron-phonon interactions. For both metals, the longitudinal vibrations of the H$_2$
leads to a decrease of the conductance at bias voltage corresponding
to the frequency of the vibration, $eV=\hbar \Omega$. In the case of
Pt electrodes, the transverse vibrations induce an increase in
conductance. This might seem surprising since the hydrogen junction supports a
single almost fully open transport channel and thus, according to the
one-channel model, inelastic scattering should always lower the
conductance. On the basis of scattering theory we showed that the
increase is a result of non-transmitting $d$-channels which couple to
the transmitting $s$-channel via the transverse modes. This is
consistent with the finding that transverse modes do not affect the
conductance in the case of Au electrodes.

We acknowledge support from the Lundbeck Foundation's Center for
Atomic-scale Materials Design and the
Danish Center for Scientific Computing.


\bibliographystyle{apsrev}

\begin{thebibliography}{44}
\bibitem{voltage} 
  Note that both the Green functions and self-energies depend on the bias voltage, however, for notational simplicity we do not show this dependence explicitly.
\end{thebibliography}

\end{document}